\documentclass[preprint2]{aastex}
\usepackage[dvips]{epsfig}
\usepackage{amsmath}

\shorttitle{Inclination Uncertainties}
\shortauthors{Barnes \& Sellwood}

\slugcomment{\today}

\newcommand{\chisq}{\ensuremath{\chi^2_r} }

\begin{document}

\title{Uncertainties in Spiral Galaxy Projection Parameters}
\author{Eric I. Barnes \& J. A. Sellwood}
\affil{Department of Physics \& Astronomy, Rutgers University, 
Piscataway, NJ 08854}
\email{barnesy@physics.rutgers.edu}
\email{sellwood@physics.rutgers.edu}

\begin{abstract}

We investigate the impact of nonaxisymmetric structure on estimates of 
galaxy inclinations and position angles.  A new minimization technique is 
used to obtain estimates of inclination and position angle from a global
fit to either photometric or kinematic data.  We discuss possible 
systematic 
uncertainties which are much larger than statistical uncertainties.  Our 
investigation reveals that systematic uncertainties associated with 
fitting photometric data dominate the formal statistical uncertainties.  
For our sample of inclined galaxies, we estimate that nonaxisymmetric 
features introduce inclination and position angle uncertainties of 
$\approx 5^{\circ}$, on average.  The magnitudes of 
these uncertainties weaken the arguments for intrinsically elliptical 
galaxy disks.

\end{abstract}

\keywords{galaxies:fundamental parameters\,---\,galaxies:photometry\,---
galaxies:kinematics and dynamics}

\section{Introduction}

We observe disk galaxies projected at random angles on the sky and must, 
therefore, determine the inclination geometry for each galaxy in order to 
deproject line-of-sight velocity data, or discover the intrinsic shape of 
the disk, etc.  The in-plane orbital speed is needed for mass estimation, 
and is also one of the parameters required for the Tully-Fisher relation 
(hereafter TFR, Tully \& Fisher 1977).  Furthermore, correction for 
internal extinction is also 
inclination-dependent (e.g., Giovanelli {\it et al.}\ 1994), affecting the 
total luminosity of the galaxy, the other parameter of the TFR.  Thus 
errors in both the inclination and position angle, propagate directly into 
dynamical mass estimates, luminosity estimates, and contribute to the 
scatter in systematic relations.

The most common procedure to determine orientation geometry is to fit 
ellipses to a photometric image of the galaxy, and to infer the 
projection geometry by assuming the disk to be intrinsically thin and 
axisymmetric.  The ellipticity of the projected image, $\epsilon=1-b/a$ 
where $b$ and $a$ are respectively the semi-minor and semi-major axis 
lengths, is related to the inclination angle through $\cos{i}=1-\epsilon$, 
for an assumed intrinsically axisymmetric, infinitesimally thin, disk.  
Here, $i$ is the inclination angle of the disk plane to the plane of the 
sky ($i=0^{\circ}$ for a face-on galaxy), and we use the symbol $\phi$ 
for position angle of the line of intersection of these two planes.  A 
slightly more complicated relation exists for galaxies that are assumed 
to be thin oblate spheroids \citep{hub26}. 
 
If 2-D velocity fields are available, they can also be used to determine 
inclinations and position angles.  The previously mentioned assumptions of 
intrinsic axisymmetry and thinness are again typical.  A drawback to this 
method is that it is poorly suited to determine inclinations where the 
rotation curve rises slowly \citep{vmw85}.  On the other hand, 2-D 
velocity fields can clearly delineate position angles.

\citet{ga91} summarize several other procedures that have been adopted to 
estimate $\epsilon$ from photometry in the presence of the complicating 
non-axisymmetric features -- spirals, bars, lop-sidedness, etc.\ -- 
manifested by virtually all disk galaxies.

The IRAF\footnote{IRAF is distributed by NOAO, which is operated by AURA, 
under contract to the NSF.} utility {\it ellipse} implements the method 
outlined by \citet{jed87} and fits ellipses to a number of isophotes 
independently, although the task can be instructed to hold any parameter, 
such as the photometric center, fixed for all.  The formal statistical 
uncertainties due to noise in the data (hereafter abbreviated to 
statistical uncertainties) in the fitted parameters are not easily 
obtained \citep{bus96}, but are generally quite small.  The user of 
this utility then has to make a judgement as to which isophote, or group 
of isophotes, offers the best estimate of the projection geometry.  A 
common prejudice is that the outer disk is more likely to be intrinsically 
round, leading to a preference for the outermost, or the average of several 
outer ellipses.  Another strategy is to identify the true shape with a 
radial range over which the fitted parameters do not vary much.

Such a procedure clearly does not utilize all the information in the 
image.  More sophisticated methods include: minimizing the $m=2$ component 
of the Fourier transform of an image \citep{gros85}, bulge-disk 
decomposition \citep{kent85}, maximizing the $m=0$ component of 
deprojected images \citep{iye82}, and fitting a parametric model for 
the disk \citep{pw00}.  We propose another global method which, like 
those just cited, assumes that the galaxy disk is infinitesimally thin 
and intrinsically axisymmetric.  Our method returns $\epsilon$ and $\phi$
values from a chi-squared minimization using a large part of the image. 
The details are summarized in the Appendix.

The statistical uncertainty of the methods described above is usually 
small, but it is clear that the possible systematic error could be large 
if the disk is intrinsically non-circular or if nonaxisymmetric features 
are present.  In this paper, we attempt to quantify the possible 
systematic error from a detailed re-analysis of the photometric and 
2-D kinematic maps of a sample of 74 spiral galaxies already reported 
by \citep{pw00} (hereafter PW).
  
After briefly introducing the data sets used in the study in \S\ref{data}, 
we give a general discussion (\S\ref{uncert}) of the statistical and 
systematic errors that can arise when estimating projection angles.  In 
\S\ref{result}, we present the results of photometric and kinematic 
fitting procedures applied to the PW galaxies.  Unlike these authors, 
however, we study possible systematic errors in estimating the projection 
parameters by comparing our fits to the kinematic and photometric data.

\section{Data} \label{data}

We have used the disk fitting methods described in the Appendix to 
estimate the projection angles, $i$ and $\phi$, from the $I$-band 
photometry and 2-D velocity fields for the 74 galaxies in the PW sample.  
PW kindly provided the reduced photometric images with the foreground 
stars removed.
 
PW also determined 2-D velocity fields for the same galaxies from H$\alpha$ 
spectroscopy taken with the Rutgers Fabry-Perot imaging spectrophotometer.
They obtained several (8 to 15) images spaced by 1 \AA 
($\approx 50$ km/s at H$\alpha$) of each galaxy from which they derived 
data cubes ($x,y,v_{los}$).  These cubes were then reduced to 2D spatial
maps by fitting the line profiles at each pixel 
with a Voigt function.  Again, PW kindly supplied us with the resulting 
velocity fields, H$\alpha$ intensities, continuum intensities, and their 
respective uncertainties.

Inclined galaxies were preferred in this sample, which is therefore 
notably deficient in nearly face-on systems.  This selection bias makes 
our study complementary to those by \citet{rz95} and \cite{and01} who 
focused on face-on systems.

\section{Statistical \& Systematic Uncertainties} \label{uncert}

Noise in the data (photometric or kinematic) creates statistical parameter 
uncertainties, $\sigma_x$, which can be measured in a variety of ways.  For
example, the curvature of the 
reduced chi-squared, \chisq, surface at the minimum, $\chisq=\chi^2_{r,{\rm 
min}}$, or better, the distance from the minimum at which 
$\chisq=\chi^2_{r,{\rm min}} + \Delta\chisq$, allows estimates of the 
statistical uncertainies in the parameters, and any covariances.  
Confidence limits can be determined from the contours at 
$\Delta \chisq = n/\nu$, where $\nu$ is the number of degrees of 
freedom, i.e., the number of pixels minus the number of free parameters 
in the fit, and $n$ is an integer.  Clearly, the magnitudes of the 
statistical uncertainties in the estimated parameters are small when many 
pixels are used, and statistical uncertainties in the projection angles 
frequently turn out to be $\ll 1^{\circ}$.

Sources of systematic uncertainties, $s_x$, in photometric ellipse fitting 
include: (1) seeing, (2) dust obscuration, (3) finite disk thickness, (4) 
non-axisymmetric features in the disk, and (5) intrinsic ellipticity of the 
disk.  Velocity field fitting is also subject to similar systematics.  
Points 4 and 5 are relevant to both methods.  Additionally, if a nearly 
edge-on galaxy with finite thickness has rotational velocities that 
decrease with height away from the midplane, fitting a thin disk model 
leads to an apparent inclination that is somewhat less edge-on than in 
reality. 

Seeing tends to bias elliptical isophotes towards becoming round.  While 
it is most severe in the central regions, the effect can be of the same 
order as an intrinsic ellipticity for moderate distances from the center 
\citep{tru01}.  Using the typical seeing value quoted in PW of 
$1''.5$ and Figure 1 of \citet{tru01}, we find that at a radius 12 times 
the seeing length a measured $\epsilon$ value is $\approx 90\%$ of its 
actual value.  This can lead to errors in $i$ of a few degrees, depending 
on the value of $i$.  However, since most of the pixels utilized to make 
the fits lie beyond 12 times the seeing length (and are less affected by 
seeing), the effect is probably less important than others we discuss below.

Since inclined galaxies suffer more extinction than do face-on galaxies, 
dust is a source of systematic error.  Its effect can be reduced by 
utilizing photometry in near IR bands where dust extinction is less 
severe, as we do in this study.  

Using simulation and analytical work, we have found that photometrically 
determined $i$ values are significantly affected by finite thickness 
only for the most edge-on galaxies (inclination $\gtrsim 80^{\circ}$).  
Photometric fits of model inclined oblate spheroids agree well with the 
formula given in \citet{hub26} relating inclination, apparent axis ratio, 
and spheroidal axis ratio.  Since the majority of the PW galaxies have 
$i \le 80^{\circ}$, we ignore this complication hereafter.

Nonaxisymmetric structures in disks, such as bars, spirals, and warps, 
are a major source of systematic error.  \citet{stock55} investigated the 
impact of spiral structure on the photometric estimation of position 
angles, particularly the case of highly inclined galaxies, finding that
spiral structure can introduce uncertainties in $\phi$ values of $\approx
5^{\circ}$.  
We refer to the impact of all such nonaxisymmetric structure on the 
photometric projection angles as the spiral effect.  

Several studies, most notably \citet{bnv81,gros85,franx92,rz95,bb99,and01},
have found hints that the disks of spirals are not perfectly axisymmetric, 
which may arise because the gravitational potential well within which the 
disk material orbits is intrinsically non-axisymmetric.  We distinguish 
this global intrinsic oval distortion of the disk from more obvious 
features in the disk, such as spiral arms and bars.  One reason for 
drawing this distinction is that the spiral effect can be investigated 
using photometric data only, while the effects of intrinsic ellipticity 
should be most clear when photometric data are compared to kinematic data 
\citep{franx92}.  Depending on how the major axis of an elliptic disk is 
projected, an intrinsic ellipticity can lead to either larger or smaller 
inferred ellipticity values, leading to over- or under-estimates of the 
inclination, and to corresponding errors in the position angle.

\section{Results} \label{result}

PW determined the projection angles for each galaxy separately from the 
photometric and kinematic data.  In general, they found somewhat different 
values from the two methods, which was inconvenient for their objective.  
They wished to compare the observed rotation curve with that predicted 
from a photometric disk mass model, which requires a common projection 
geometry for both.  They argued that the inclination and position of the 
geometric center could be determined most reliably from the photometry, 
while the position angle of the major axis was best determined from the 
kinematics.  They then re-fitted both data sets with the preferred 
parameter from the other set held fixed.

Our inclinations derived from photometry are in good agreement with 
the values reported in PW.  While reassuring, the agreement is scarcely 
surprising as we are using similar methods to fit the same photometric 
data.  Discrepancies arise, in part, because we have fitted for both 
$i$ and $\phi$ while in PW $\phi$ was held fixed at the value determined 
from the kinematics, but the good agreement (the average difference 
between our $i_{\rm phot}$ values and those of PW is $2.9^{\circ}$
shows that this constraint has a small effect.  The largest difference
is $19^{\circ}$ and occurs for ESO 510G11, a galaxy with very strong
spirals that extend into the outer parts of the disk.

\begin{figure}
\plotone{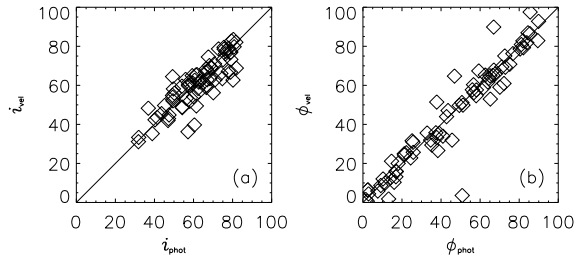}
\caption{a) Inclinations determined from photometric data ($i_{\rm phot}$) 
compared with those determined from kinematic data ($i_{\rm vel}$).  b)
Position angles determined from photometric data ($\phi_{\rm phot}$)
compared to those determined from kinematic data ($\phi_{\rm vel}$).  The
extreme outlier is discussed in the text.  The solid lines in both
frames are lines of equality.
\label{comp}}
\end{figure}

One of our principal objectives here is to understand the discrepancy 
between the photometric and kinematic values of the projection angles for 
each galaxy.  The magnitudes of these differences are illustrated in 
Figures \ref{comp}a and \ref{comp}b.  The statistical uncertainties 
are much smaller than the symbols in both figures.

The inclinations show considerable spread around perfect agreement (Figure 
\ref{comp}a).  The average photometric/kinematic inclination difference 
is $\approx 5^{\circ}$ with a standard deviation of similar magnitude.  
A comparison of position angles 
(Figure \ref{comp}b) shows a similar spread, with one extreme outlier 
which we discuss briefly below.  The average difference between position 
angle estimates is $\approx 4^{\circ}$ (excluding the extreme outlier).
Note, however, that this galaxy sample is deficient in nearly face-on 
galaxies and that \citet{and01} found larger differences in a sample of 10
low inclination ($i \lesssim 20^{\circ}$) galaxies.

\begin{figure}
\plotone{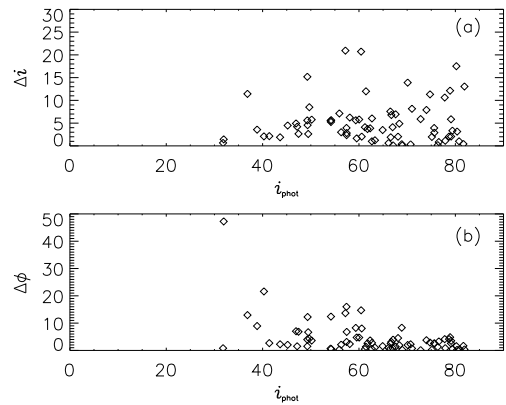}
\caption{a) The difference between inclination angles determined from
photometric data and kinematic data plotted against the photometric 
inclination.  b) The difference between position angles determined 
photometrically and kinematically versus the photometric inclination.  
Note that the $\Delta \phi$ scatter shrinks as the inclination increases.  
The extreme outlier is discussed in the text.
\label{diffvsi}}
\end{figure}

Figures \ref{diffvsi}a and \ref{diffvsi}b show the relations between 
the 
inclination and position angle differences and the photometric inclination 
of the galaxy.  While there is significant spread in Figure 
\ref{diffvsi}b, there appears to be a trend for more highly 
inclined galaxies to have smaller position angle differences.  Such a 
trend is expected, since any nonaxisymmetric structures present in 
highly-inclined galaxies should have only a small effect on 
$\phi_{\rm phot}$.  This trend also agrees with the aforementioned larger 
differences found for low inclination galaxies \citep{and01}.

The largest position angle difference in Figure \ref{diffvsi}b 
corresponds 
to a galaxy which has a low inclination and strong spirals which extend 
to the outer reaches of the disk.  It is a prime example of the kind of 
systematic error that can be attributed to the spiral effect.  
Additionally, this kind of misalignment could cause serious problems if 
single-slit spectroscopic data were to be utilized to determine rotation 
curves.  Misalignment between the slit and the true major axis results 
in an underestimate of rotational speed and hence dynamical mass.

\subsection{Fixed Centers vs.\ Floating Centers} \label{centers}

\begin{figure}
\plotone{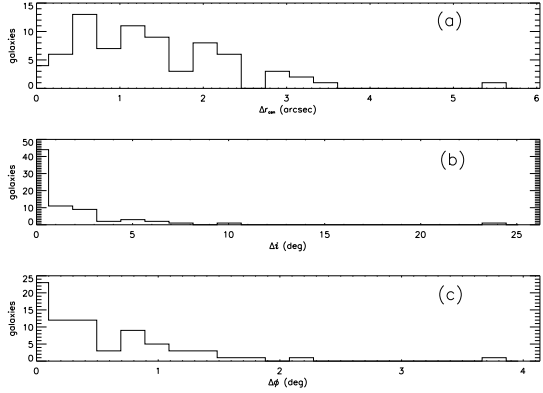}
\caption{a) Histogram of the distance between the photometric center and 
the fitted kinematic center.  The difference is $<2.\arcsec7$ for the 
majority (90\%) of the galaxies.
b) Histogram of the change in estimated inclination between the 
fixed-center and floating-center fits.  95\% of the galaxies have 
$\Delta i<7^{\circ}$.
c) Histogram of the change in position angle between the fixed-center and 
floating-center fits.  None of the galaxies has $\Delta \phi > 4^{\circ}$. 
\label{hists}}
\end{figure}

In all the work reported elsewhere in this paper, we have required the 
rotation center of our fitted model to coincide with the peak of continuum 
brightness in the image (see Appendix).  In this subsection, we report the
consequences of relaxing this requirement and determine the kinematic 
center from the velocity field alone.  The offset between the kinematic 
and photometric centers, and the resulting differences in
estimates for $i$ and $\phi$ are illustrated in Figure \ref{hists}.

Figure \ref{hists}a shows that the kinematic center is 
$\lesssim 2.\arcsec7$ from the photometric center in the majority 
(67 of 74) of galaxies.  The galaxy for which this distance is largest,
ESO 322G42, has a slowly rising rotation curve, which creates nearly 
parallel isovelocity contours; the location of the center and the overall 
systemic velocity are poorly constrained by such data, but the inner slope 
of the rotation curve is independent of the center location.  Six other 
galaxies are in the tail ($\Delta r_{\rm cen} > 2.\arcsec7$):  One (ESO 
445G19) has a strongly asymmetric inner velocity field, while the others 
are either nearly edge-on and have roughly parallel isovels and slowly
rising rotation curves, or lack H$\alpha$ emission near the center; the 
location of the rotation center is not tightly constrained in the absence 
of kinematic information near the center.

As evidenced by the histograms in Figures \ref{hists}b and c, the 
projection angles that result from fitting with the kinematic center free 
to move agree with those from the fixed-center fits for the majority of 
the galaxies.  In fact, none of the $\phi$ values changes by more than 
$4^{\circ}$.  The changes in inclination show a larger range, but 70 of 
74 galaxies have $\Delta i < 7^{\circ}$.  The galaxy with the largest 
$\Delta i$ shows evidence of being tidally distorted, having a long 
filament of anomalous velocities extending from its edge.  The two 
galaxies with the next largest $\Delta i$ values both have strong spirals 
resulting in velocity fields with strong distortions.  The remaining 
galaxy in the tail of the $\Delta i$ histogram is a highly inclined galaxy 
that is also in the tail of the $\Delta r_{\rm cen}$ histogram.

We find that unless the kinematic center is poorly constrained by a slowly 
rising rotation curve, it lies within $\approx 2.\arcsec5$ of the position 
of the photometric center for most galaxies.  So, if rotation curves 
of our galaxy sample were determined from single-slit spectra taken 
through the 
point of maximum continuum, they would not be significantly biased towards 
rising slowly.  Allowing a small shift in the kinematic center generally 
makes little difference to the estimated projection angles, except in a 
few special cases.

\subsection{Spiral Effect Systematics}\label{stock}

In an attempt to estimate the magnitude of the uncertainties due to the 
spiral effect, we have proceeded as follows.  (Note that these values must 
be viewed as lower limits to the true systematics, which are unknown.)
We use a first estimate of the photometric projection parameters from our 
\chisq minimization procedure 
to deproject a galaxy image.  We then rotate it through an angle $\beta$, 
so that the nonaxisymmetric structure lies in a new orientation, and then 
reproject it to the original best-fit $i$ and $\phi$.   We then apply our 
photometric fitting apparatus to this new image to find new best-fit 
$i$ and $\phi$ values.  We interpret the spread in both $i$ and $\phi$ as 
$\beta$ varies through a half-rotation as the measures of systematic 
uncertainties $s_i$ and $s_{\phi}$ due to the presence of nonaxisymmetric 
features in the image.  A few representative plots of $i$ and $\phi$ versus
$\beta$ are presented in the next section.

\begin{figure}
\plotone{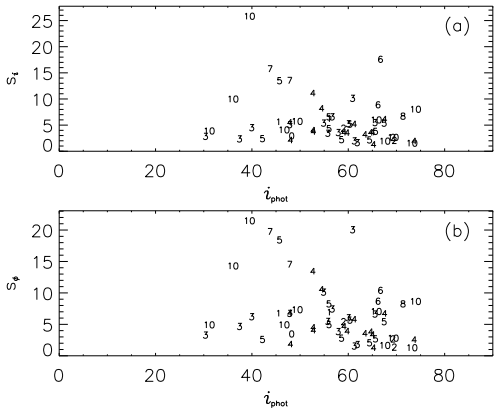}
\caption{Systematic uncertainties in (a) photometric inclination, (b) 
photometric position angle.  The numbers denote the $T$-type of the 
galaxy.  Note that the scatter decreases with increasing $i_{\rm phot}$.
\label{sysfig}}
\end{figure}

The results of this investigation are shown in Figure \ref{sysfig}.  
Galaxies
that have $i \gtrsim 75^{\circ}$ have been excluded from this analysis 
because deprojection must be considered suspect for such large inclination 
angles.  The general trend of decreasing $s_i$ and 
$s_{\phi}$ with increasing $i$ concurs with our earlier assertion that 
nonaxisymmetric structure should compromise the model fits more modestly 
for highly inclined galaxies.  The galaxy with the highest values of $s_i$
and $s_{\phi}$ (ESO 323G39) appears to have a significant bar.

The numbers in Figure \ref{sysfig} denote different galaxy $T$-types, where
low numbers indicate early-type spirals and high numbers indicate 
late-type spirals and irregulars.  The different $T$-types do not nicely 
segregate in either frame of Figure \ref{sysfig}.  However, there is a 
statistical trend for later type galaxies to have larger $s_i$ and 
$s_{\phi}$ values.  Early type spirals ($T$-types 0-3) have average 
$s_i$ values of $\approx 3^{\circ}$ and average $s_{\phi}$ values of 
$\approx 4^{\circ}$.  Spirals with 
$T$-types between 4 and 6 have average $s_i \approx 4^{\circ}$ and 
$s_{\phi} \approx 4^{\circ}$.  Late type spirals ($T$-types $\geq 7$) have 
average $s_i \approx 6^{\circ}$ and $s_{\phi} \approx 7^{\circ}$.  The
differences between the means of the early and intermediate spirals are
not statistically significant.  However, the differences between the early
and late types' means are significant at the 95\% level.  For all 
galaxies with $i<75^{\circ}$, the average $s_i \approx 4^{\circ}$ and 
$s_{\phi} \approx 5^{\circ}$, with 
a large spread for both values.  The magnitudes of $s_{i}$ and $s_{\phi}$ 
suggest that the uncertainties in $i$ and $\phi$ values derived from 
photometry are dominated by systematic uncertainty due to nonaxisymmetric 
structure.

\subsection{Intrinsic Ellipticities vs.\ the Spiral Effect}\label{intell}

We find differences between the projection angles determined by fitting 
photometric and kinematic data.  \citet{franx92}, \citet{and01}, and 
others suggest that such disagreements are 
evidence of intrinsically elliptical disks, but such a conclusion requires 
that the projection angles can be determined with sufficient precision for 
their differences to have meaning.

Our investigation of the spiral effect points to significant systematic 
errors in the photometric determination of projection angles which are of 
the same magnitude as the errors suggested by \citet{franx92} to explain 
the scatter in the TFR.  This leads us to wonder whether differences 
between photometric and kinematic projection angles are evidence of 
anything more than our ignorance of the true angles.

Fortunately, we can use our 2D velocity fields to search for evidence of 
intrinsic ellipticity.  \citet{fgdz94} formulate elliptical orbits in 
disks in the epicyclic approximation with a small $m=2$ forcing term added 
to the central attaction.  This distortion is entirely contained within 
the plane of the disk.  For the special case of a flat rotation curve, the 
distortions in the radial and tangential directions are equal and 
can be described by,
\begin{subequations}
\begin{equation}
v_{\rm R}=v_{\rm circ} \epsilon_{\rm int} \sin{2(\psi-\gamma)},
\end{equation}
and
\begin{equation}
v_{\phi}=v_{\rm circ}[1+\epsilon_{\rm int} \cos{2(\psi-\gamma)}].
\end{equation}
\end{subequations}
Here $v_{\rm circ}$ is the circular speed, $\epsilon_{\rm int}$ is the 
intrinsic ellipticity of the disk, $\psi$ is the angle in the disk 
measured from the projected major axis, and $\gamma$ is the angle between 
the projected major axis and the elliptical disk minor axis.  When such a 
disk is inclined, the line-of-sight velocity field appears as that of an 
axisymmetric disk with small correction terms added,
\begin{eqnarray} \label{evlos}
\lefteqn{v_{\rm los,ellip} = } \nonumber \\
 & & v_{\rm los,axisym} - \nonumber \\
 & & \epsilon_{\rm int} v_{\rm circ} \sin{i}
(\cos{2\gamma} \cos{\psi} - \sin{2\gamma} \sin{\psi}),
\end{eqnarray}
where $v_{\rm los,axisym}$ is the line-of-sight velocity for an 
axisymmetric disk with the same projection angles.  Equation \ref{evlos} 
demonstrates that a mildly elliptical disk creates an $m=1$ 
distortion in the residual velocity field, at a position angle which 
depends on $\gamma$.  (There is also an $m=3$ term in the more general 
case of a non-flat rotation curve, see Franx, van Gorkom, \& de Zeeuw 1994
for details.)

In order to quantify the strength of such distortions, we introduce two 
quantities, $P$ and $Q$ which we derive from the velocity residuals in the 
following way.  We divide the residual field inside the outermost fitting 
ellipse into 15 elliptical annuli, and divide each annulus into quadrants.  
We label the first quadrant (I) to be that bounded by the receding side of 
the projected major axis and the minor axis counter-clockwise from it, and 
the remaining quadrants (II, III, IV) are in the usual order.  For each 
quadrant in each annulus, we calculate the average of the residuals 
$\bar{r}_{n,{\rm quad}}$, normalized by the average of the magnitude of 
the residuals in the entire field.  We then define $P$ and $Q$ values for 
each annulus as follows,
\begin{subequations}
\begin{equation}
P=| \bar{r}_{n,{\rm I}} + \bar{r}_{n,{\rm II}} - 
    \bar{r}_{n,{\rm III}} + \bar{r}_{n,{\rm IV}} |
\end{equation}
and
\begin{equation}
Q=| \bar{r}_{n,{\rm I}} + \bar{r}_{n,{\rm IV}} - 
    \bar{r}_{n,{\rm II}} + \bar{r}_{n,{\rm III}} |.
\end{equation}
\end{subequations}
Basically, $P$ measures the strength of an $m=1$ distortion across the 
major axis, while $Q$ measures the same thing across the minor axis.  
Plotting $A \equiv \sqrt{P^2+Q^2}$ values versus the semi-major axis of 
the annulus provides an asymmetry profile.

Using equation \ref{evlos}, we have constructed a model residual velocity 
field for a disk with nested similar concentric elliptical orbits.  The 
resulting asymmetry profile rises rapidly in the central region of the 
residual field, but then flattens and remains constant for the remaining 
radial range.  None of the galaxies we have studied show this kind of 
asymmetry profile.  Galaxies which have bars and/or strong spirals show 
peaks in their asymmetry profiles (for example, Figures \ref{439g20}c VI 
and \ref{439g20}d VI), but the distortion is localized and not seen 
throughout the entire disk.  It is also interesting to note that a 
higher-than-average level of asymmetry in a velocity field is not always 
associated with a significant difference between photometric and 
kinematic projection angles.

\section{Individual Galaxies}\label{gbu}

Here we present several individual galaxies that illustrate the difficulty 
of measuring accurate projection angles.  For each galaxy, we show an 
$I$-band image, a 2D velocity field, a residual photometric image, a 
residual velocity field, a plot illustrating the spiral effect, and an 
asymmetry profile.  The photometric images are presented in grayscales with
units of magnitudes per arcsecond squared.  The white circles in the 
photometric images mark the locations of stars that have been removed.  
In the 
photometric residual images, the central white circles mask the pixels that 
are associated with a bulge and/or bar and are therefore not included in
the fits.  The velocity fields are given in 
units of kilometers per second.  The velocity and residual velocity fields 
display the pixels with measurement uncertainties less than 25 
km/s.  The suppressed pixels do not strongly influence the fitted 
parameters.  The spiral effect plot is composed of one frame showing the 
response of the fitted $i$ value to the rotation of the image and another 
showing how $\phi$ changes with the rotation.  In the asymmetry profiles, 
asterisks mark the $A$ values for the residual velocity fields.

Since photometric data is often more extensive than kinematic data, some
of the images in the following figures may appear unequal.  As described
in the Appendix, the data used by our fitting routine is different for
photometric and kinematic fits.  However, we utilize the maximum amount
of information in both cases.  To test the impact of the differing
radial extents of the data, we fit the photometric data over the same 
radial range as the kinematic data.  There were no significant changes
in the average differences between photometric and kinematic inclinations
or position angles.  For the remainder of the paper, the results given
are derived using the maximum amount of information for the data set.

\subsection{ESO 439G20 \& ESO 322G45}

The photometric images of ESO 439G20 (Figure \ref{439g20}a I) and 
ESO 322G45 (Figure \ref{322g45}b I) appear to be free of strong bars and 
spiral structure.  Their respective velocity fields also appear to be 
those of simple inclined axisymmetric disks (Figures \ref{439g20}a II and 
\ref{322g45}b II).  The naive assumption is that these two ``nice'' 
galaxies should show little asymmetry and should have good agreement 
between their photometric and kinematic projection angles.

For ESO 439G20, the photometric and kinematic $i$ difference 
is $3.5^{\circ}$ and the $\phi$ difference is $1.6^{\circ}$.  The lack of 
strong nonaxisymmetric structure is evident from the photometric residual 
(Figure \ref{439g20}a III) and the plot illustrating the spiral effect 
(Figure \ref{439g20}a V).  The asymmetry profile (Figure \ref{439g20}a VI) 
shows little influence of bars or spirals.  The last $A$ value is due to 
the small blue and red spots on either side of the minor axis on the right 
side of Figure \ref{439g20}a IV.

ESO 322G45 has an $i$ difference of $4.7^{\circ}$ and a $\phi$ 
difference of $4.5^{\circ}$.  Figures \ref{322g45}b III and \ref{322g45}b V 
are evidence that there is little asymmetry in the photometric image.  
Again, the asymmetry profile (Figure \ref{322g45}b VI) is fairly 
featureless.  The last two $A$ values may be the effect of a weak spiral 
arm that appears in the photometric residual image $\approx 20$ arcsec 
from the center.  It is surprising that such a featureless galaxy has 
large $i$ and $\phi$ differences, but they cannot be symptoms of intrinsic 
ellipticity in the disk, for the following reasons:  An elliptical disk 
viewed in projection should generally give rise to the type of asymmetry 
profile discussed in \S \ref{intell}, which is not seen in Fig. 5b VI.
The special case of projection about one of the disk's principal axes 
could mask this signature, but then we should not expect a significant
$\Delta\phi$, which we also find.  Thus, we are confident that the
differences in this case are not due to an intrinsically elliptic disk.

These two galaxies serve to illustrate the point that just because a 
galaxy's image or velocity field appears smooth or simple does not imply 
accurate estimation of projection angles.

\subsection{ESO 267G29 \& ESO 569G17}

In contrast to the previous examples, we present two galaxies whose 
photometric images (Figures \ref{267g29}c I and \ref{569g17}d I) show 
strong bars and spiral structure.  Their velocity fields 
(Figures \ref{267g29}c II and \ref{569g17}d II) contain distorted 
isovelocity contours, evidence of strong nonaxisymmetric structure.  In 
these cases, it is not unreasonable to believe that estimation of 
projection angles will be problematic.

ESO 267G29 has a photometric/kinematic $i$ difference of $20^{\circ}$ and a 
$\phi$ difference of $15^{\circ}$.  The strong open spiral can be clearly 
seen in Figure \ref{267g29}c III and the effect of the central bar is 
obvious from the $m=1$ residual pattern in central region of Figure 
\ref{267g29}c IV.  The asymmetry profile (Figure \ref{267g29}c VI) clearly 
shows the effect of the central bar with the rise in $A$ at large radius 
due most likely to the strong spiral also present.  

To again contrast, the differences between photometric and kinematic $i$ 
and $\phi$ values for ESO 569G17 are $< 1^{\circ}$.  A dusty multi-armed 
spiral pattern can be seen in Figure \ref{569g17}d III.  As before, the 
effect of the central bar is evident in the residual velocities (Figure 
\ref{569g17}d IV).  Like ESO 267G29, the asymmetry profile in Figure 
\ref{569g17}d VI shows the impact of a central bar and some effect from 
spiral arms.  

It is not surprising that if the central region of the velocity field of 
ESO 267G29 is ignored, as it is in the photometry, the differences between 
photometric and kinematic $i$ and $\phi$ drop to $\approx 10^{\circ}$.  In 
contrast, ignoring the central region of the velocity field of ESO 569G17 
increases the discrepancy between photometric and kinematic projection 
angles by $\approx 1^{\circ}$.  So, even though photometric images and 
kinematic maps are complicated, the projection angles derived from them 
can be consistent.  These two examples also suggest that the details of 
the spiral structure present can significantly alter the 
photometric/kinematic discrepancy.

\subsection{ESO 215G39}

As our final example, we present a ``typical'' spiral galaxy.  The 
photometric/kinematic $i$ difference is $1.9^{\circ}$ and the $\phi$ 
difference is $2.3^{\circ}$.  The photometry and residuals (Figures 
\ref{215g39}e I and \ref{215g39}e II) show moderate spiral structure and 
the velocity field (Figure \ref{215g39}e II) is fairly simple.  The 
velocity residual field (Figure \ref{215g39}e IV) shows some evidence of 
elliptical orbits in the central region.  Again, the asymmetry profile 
(Figure \ref{215g39}e VI) contains evidence of either a weak central bar 
or spirals near the center.  Interestingly, the magnitudes of these 
photometric/kinematic differences are the same as the predicted
differences due to the spiral effect for this galaxy.

\section{Discussion \& Conclusions} \label{concl}

We have developed and tested a global fitting method with a novel
minimization technique to obtain disk galaxy 
inclinations and position angles from either photometric or kinematic data.
Since we fit a thin axisymmetric disk model to all the data at once, we 
determine globally optimal values for the geometric parameters, 
$\epsilon$ and $\phi$.  This contrasts with the IRAF {\it ellipse} utility 
which returns independent $\epsilon$ and $\phi$ values for each of many 
ellipses, and cannot be constrained to find the best common values.

We map the \chisq surface to find the best-fit inclination and position 
angle and to determine the statistical uncertainty for each fit.  Due to 
the large number of data points (typically $\gtrsim 10^3$), the statistical
uncertainty is very small.  The $i$ and $\phi$ values determined using our 
method differ only slightly from those previously estimated for the same 74
spiral galaxies by \citet{pw00}.  While we have held the rotation center 
fixed at the location of brightest continuum, changes in derived quantities
are generally small when we allow the kinematic center to move from this 
point.
 
A major result of this study is that systematics dominate photometrically 
determined inclination and position angle uncertainties.  We estimate that 
systematic uncertainties due to nonaxisymmetric structure such as spirals
and/or bars are, on average, $\approx 4^{\circ}$ for inclination estimates 
and $\approx 5^{\circ}$ for position angle estimates for galaxies in this
sample.  Again, we point out that the complete systematic error is unknown,
and these estimates must be taken as lower limits.  It is 
reassuring that the position angle uncertainties tend to
be smaller for more edge-on galaxies and that later type spirals generally 
have larger systematic inclination and position angle uncertainties.  It 
seems that the only way to reduce these systematic errors would be to 
model the nonaxisymmetric structure in each galaxy.

It is interesting to note that the average difference in inclination and 
position angle derived from the photometric and kinematic data turn out to 
be roughly the same size as our estimates for uncertainties due to 
nonaxisymmetric structure.  Since these differences are also similar to 
those expected from intrinsically elliptical disks, they reinforce the 
concern raised by \citet{franx92} and \citet{rz95} that spiral structure 
can be confused with intrinsic disk ellipticity.  As discussed in \S 
\ref{gbu}, the (non)existence of spirals and/or bars does not necessarily
imply (small) large differences between photometric and kinematic 
projection angles.  We also find no clear 
evidence for the velocity field distortions expected for intrinsically 
elliptic disks. 

To conclude, we have found a difference between galaxy projection angles 
determined from photometric and kinematic data for 74 galaxies.  We have 
also found that nonaxisymmetric structures, such as spirals and bars, 
impact the galaxies' photometric projection angles to about the same 
extent as the differences between photometric and kinematic projection 
angles.  Further, we see no direct evidence for intrinisically elliptical
disks in any of our galaxies' residual velocity fields.  In light of these 
points, it seems unnecessary to explain photometric/kinematic projection 
angle differences with intrinsic disk ellipticities.

\section*{Acknowledgments}
We are indebted to Povilas Palunas for providing us with the reduced data 
and to our colleagues Tad Pryor, Ted Williams, and Arthur Kosowsky for 
much thoughtful advice.  Thanks also to Matt Bershady and the anonymous 
referee for helpful suggestions to improve the paper.  This work was 
supported by NASA SARA grant NAG 5-10110.

\appendix

\section{Procedure to Estimate Projection Angles} \label{proc}

The principle of our method is to use as much information as possible from 
a photometric or kinematic image to estimate the two angles of projection.
If a galaxy were a razor-thin, axisymmetric disk and the orbits were 
circular, then the surface brightness or in-plane orbital speed would be 
constant on a nested set of concentric, co-axial, similar ellipses.  We 
utilize the entire velocity field, from the center of the galaxy to beyond 
its edge, but our photometric fits are to only those parts of the image not
strongly contaminated by bulge and/or bar components.

Our input data are a set of data values $\{D_i\}$, and their associated 
uncertainties, $\{\sigma_i\}$, for each of the $N$ pixels in the image.  
For photometric data, on the one hand, the $\{D_i\}$ are pixel intensity 
values, for which the principal source of noise is photon counting.  We 
estimate the sky brightness from part of the image away from the galaxy 
and subtract its value before beginning the fit.  For kinematic data, on 
the other hand, the $\{D_i\}$ are line-of-sight velocity values and the
associated uncertainties result from the fitting of the Fabry-Perot data
cube.

We choose the center to have the coordinates $(x_c,y_c)$ of the pixel that 
has the largest intensity for a photometric image, or the largest 
continuum intensity for the velocity map.  We generally hold these 
coordinates fixed for the entire fitting procedure, although our method 
can allow the center to move if desired.

We fit a non-parametric profile that is characterized by a tabulated set of 
values $\{M_k\}$, for a number $K$ of semi-major axes.  Each $M_k$ has the 
same value around an ellipse, and all ellipses have a common center and 
ellipticity, $\epsilon$, and major-axis position angle, $\phi$, embodying 
our assumption that the disk is thin and axisymmetric.  For photometric 
data, the $\{M_k\}$ are isophote intensities, while they are circular 
velocities for a kinematic map.  Note such a tabular form makes no 
assmuption about the radial profile of either the light or the 
orbital speed.  We derive a model prediction at a general location in the 
image by finding the semi-major axis of the ellipse that passes through 
the point in question, and then interpolate for the predicted pixel 
value from the $\{M_k\}$.

For any values of the projection parameters ($\epsilon,\phi$), we form
\begin{equation}\label{chi2}
\chisq=\frac{1}{\nu} \sum^N_{i=1} \left[ \frac{D_i - 
\sum^K_{k=1}w_{k,i}M_k}{\sigma_i} \right] ^2,
\end{equation}
where the $w_{k,i}$ are the weight factors that describe the interpolation 
scheme, and $\nu$ is the number of degrees of freedom in the fit.  Since 
\chisq is linear in the $M_k$s, the $M_k$s that minimize \chisq can be 
determined by setting the derivative of eq.~(\ref{chi2}) with respect to 
each $M_j$ equal to zero and solving the resulting set simultaneous 
equations, 
\begin{equation} \label{ikeq}
\sum^K_{k=1}\left( \sum^N_{i=1} \frac{w_{j,i}}{\sigma_i} 
\frac{w_{k,i}}{\sigma_i}\right) M_k= 
\sum^N_{i=1}\frac{w_{j,i}}{\sigma^2_i} D_i.
\end{equation}
The resulting $M_k$ values are used to determine \chisq.  Our 
implementation uses linear interpolation to determine the $w_{k,i}$ 
factors, but more complicated schemes would be straightforward to 
implement.  Any 
pixels that lie outside the final fitting ellipse have their model values 
linearly extrapolated from the values of the last two ellipses and their 
weights smoothly decreased depending on their distance from the last 
ellipse.  Such pixels have little impact on the best-fit parameters.

We use slightly different methods to determine which pixels will be used 
in the fitting procedure for the photometric and kinematic data.  In both 
cases, we create a pixel list only once which is used throughout the 
fitting procedure.  With photometric data, we exclude pixels contaminated 
by foreground stars or cosmic rays.  We also exclude those pixels inside a 
circular aperture containing a central bulge or bar, and set the smallest 
semi-major axis of the $K$ ellipses equal to the radius of this inner 
aperture.  The outer limit of the pixel list is defined by an outer 
elliptical aperture with semi-major axis 10\% larger than that of the 
outermost fitting ellipse, using an initial estimate of the ellipticity.  
The pixel list for the kinematic data contains all pixels that have 
continuum intensity values $\geq 0.5\%$ the central continuum intensity.  
We add, in quadrature, a fixed velocity dispersion of $7\;$km~s$^{-1}$ to 
the estimated velocity uncertainty in the line fit.
Some anomalous velocities remain in the velocity map, 
possibly caused by intervening gas not associated with the galaxy of 
interest or from Voigt profile fits near the velocity limits of the 
Fabry-Perot data cube; we eliminate such values by applying a simple cut 
to exclude pixels with velocities more than 500 km/s away from the data
value at the fixed center.

We vary the two projection angles, $\epsilon$ and $\phi$, recomputing the 
$\{M_k\}$ at each iteration to determine their values that minimize \chisq. 
We use a downhill simplex method, specifically the {\it amoeba} routine 
described in \citet{press92}, which requires only \chisq function 
evaluations (no derivatives) and initial estimates for $\epsilon$ and 
$\phi$.  Following the recommendation of \citet{press92}, the $\epsilon$
and $\phi$ 
estimates returned are re-used as initial guesses for up to 5 subsequent 
calls to {\it amoeba} to converge to the \chisq minimum.  We have verified 
the minimum location by inspecting the \chisq surface surrounding the 
best-fit values.  Projections of the contour where 
$\chisq=\chi^2_{r,{\rm min}}+1.0/\nu$ onto the $\epsilon$ and $\phi$ axes 
provide 1-$\sigma$ statistical uncertainty measurements.

Figure \ref{iplot} shows the results of photometric fitting for one galaxy 
in our sample (ESO 215G39).  The model isointensity values are shown as 
filled circles superimposed on data values.  The solid lines encompass 
68\% of the data values.  The statistical uncertainties of the parameters 
(dot values) are smaller than the symbols.  The rotation curve with 
1-$\sigma$ parameter uncertainties resulting from the kinematic fit for 
the same galaxy is shown in Figure \ref{vplot}.

\begin{figure}
\caption{(I) $I$-band image of the galaxy.  The white circles show the 
locations of removed foreground stars.  The units are 
${\rm mag}/{\rm arcsec}^2$.  (II) The line-of-sight velocity field from 
Fabry-Perot interferometry.  The values listed are heliocentric 
velocities in km/s.  (III) Photometric residuals using 
the best-fit $i_{\rm phot}$ and $\phi_{\rm phot}$.  The units are ${\rm 
mag}/{\rm arcsec}^2$.  The ellipse marks the location of the last 
ellipse used in the fitting routine.  The central white circle masks the 
pixels that are associated with a bulge and/or bar.  (IV) The residual 
velocity field using the best-fit $i_{\rm vel}$ and $\phi_{\rm vel}$.  As 
explained in the text (see \S \ref{gbu}), only those pixels with 
measurement uncertainties $<25\;$km/s are shown.
(V) The variation of the fitted $i$ and $\phi$ as the image is rotated as 
described in \S \ref{stock}  (VI) The radial variation of $A$ to reveal the 
asymmetry profile.  Discussion for each galaxy is given in the text.
\label{439g20}}
\end{figure}

\addtocounter{figure}{-1}
\begin{figure}
\caption{Continued
\label{322g45}}
\end{figure}

\addtocounter{figure}{-1}
\begin{figure}
\caption{Continued
\label{267g29}}
\end{figure}

\addtocounter{figure}{-1}
\begin{figure}
\caption{Continued
\label{569g17}}
\end{figure}

\addtocounter{figure}{-1}
\begin{figure}
\caption{Continued
\label{215g39}}
\end{figure}

\begin{figure}
\plotone{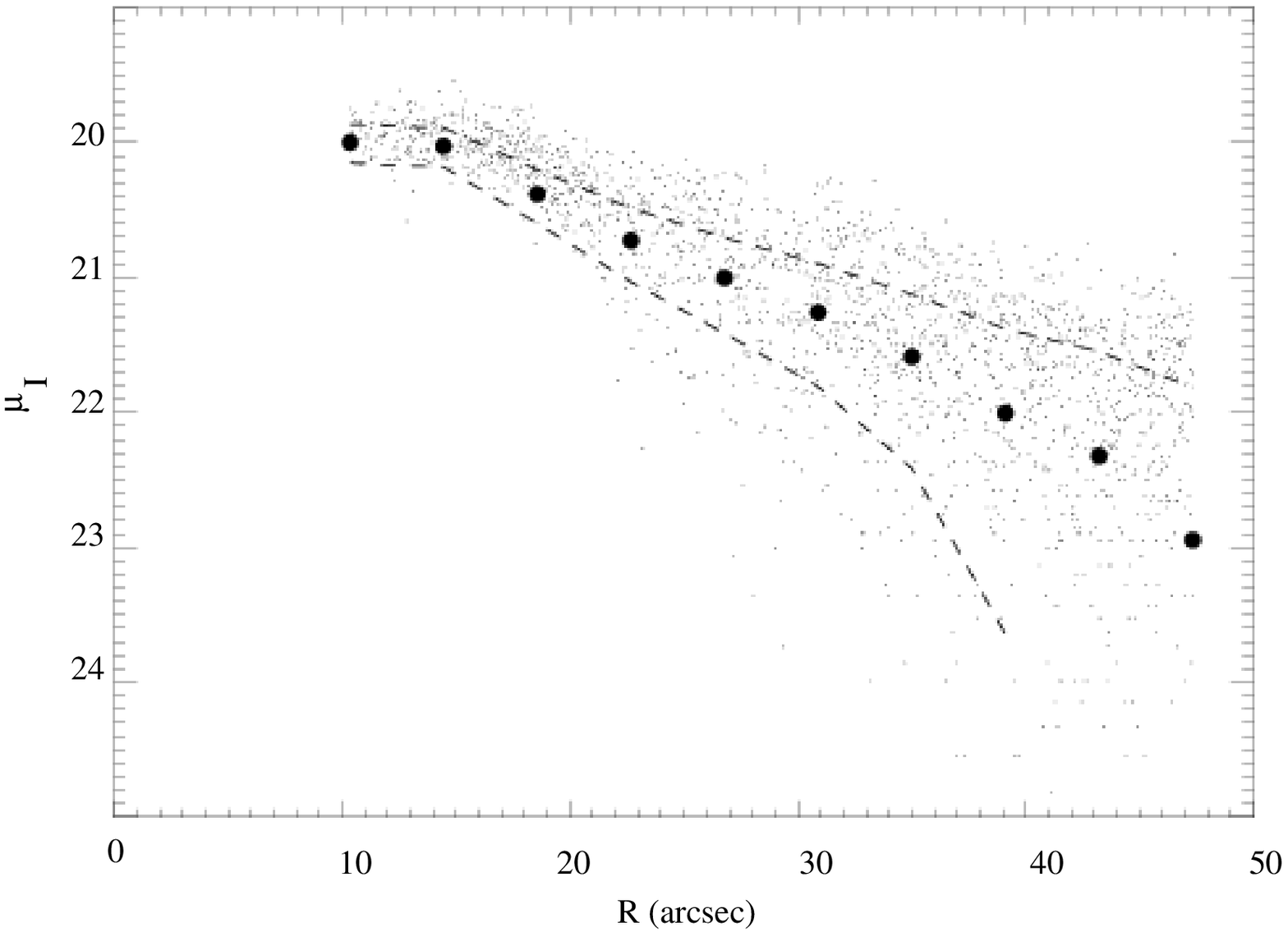}
\caption{Model isophote values $\{M_k\}$ (filled circles) versus 
semi-major axis for the galaxy ESO 215g39.  Some indicative (but not all)
sky-subtracted data values are plotted.  The dashed lines contain 68\% of 
the data values and the 1-$\sigma$ uncertainties for the model values are 
much smaller than the filled circles.
\label{iplot}}
\end{figure}

\begin{figure}
\plotone{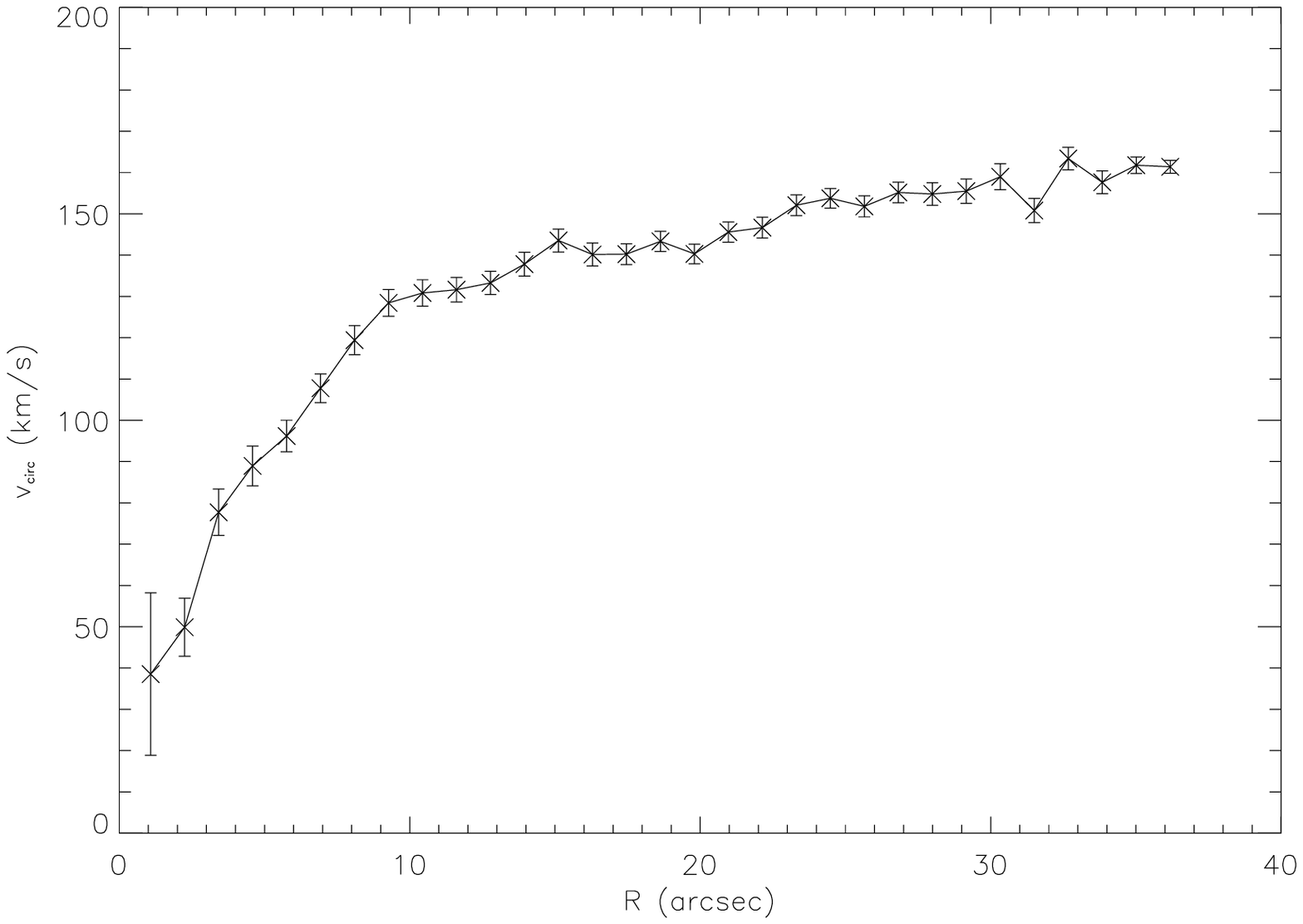}
\caption{Estimated rotation curve for the galaxy ESO 215g39.  The crosses 
with error bars are the $\{M_k\}$ values returned from the kinematic 
fitting routine together with their 1-$\sigma$ statistical uncertainties.
\label{vplot}}
\end{figure}

\end{document}